# Title

All-Dielectric Nanolaser with Subnanometer Field Confinement

## Authors

Hao Wu,[1] Peizhen Xu,[1] Xin Guo,[1*] Pan Wang,[1] Limin Tong[1,2*]

## Affiliations

[1] Interdisciplinary Center for Quantum Information, State Key Laboratory of Modern Optical Instrumentation, College of Optical Science and Engineering, Zhejiang University, Hangzhou 310027, China.

[2] Collaborative Innovation Center of Extreme Optics, Shanxi University, Taiyuan 030006, China.
[*]guoxin@zju.edu.cn; phytong@zju.edu.cn.

## Abstract

The ability to generate a laser field with ultratight spatial confinement is important for pushing the limit of optical science and technology. Although plasmon lasers can offer sub-diffraction-confined laser fields, it is restricted by a trade-off between optical confinement and loss of free-electron oscillation in metals. We show here an all-dielectric nanolaser that can circumvent the confinement-loss trade-off, and offer a field confinement down to subnanometer level. The ultra-confined lasing field, supported by low-loss oscillation of polarized bound electrons surrounding a 1-nm-width slit in a coupled CdS nanowire pair, is manifested as the dominant peak of a $TE_0$-like lasing mode around 520-nm wavelength, with a field confinement down to 0.23 nm and a peak-to-background ratio of ~32.7 dB. This laser may pave a way towards new regions for nanolasers and light-matter interaction.

# MAIN TEXT

## Introduction

Lasers with tighter field confinement is a key to lower-dimensional light-matter interaction for applications ranging from optical microscopy, sensing, photolithography to information technology (*1–3*). Since the early age of laser technology, pursuing tighter field confinement in a lasing cavity mode has been a continuous effort. Generally, limited by optical diffraction, a dielectric lasing cavity is unable to confine fields much better than half the vacuum wavelength ($\lambda_0/2$) (*2*, *4*). The emerging plasmonic nanocavity opens a route towards deep-sub-diffraction lasing field with optical confinement down to $\lambda_0/30$ (*5–11*), and has shown great potential in applications such as photonic integrated interconnects and sensing (*3*, *11*). However, due to the trade-off between optical confinement and plasmon loss of oscillating free electrons that is intrinsically originated from Landau damping in metals (*12*), scaling down the cavity size of a plasmon nanolaser is a great challenge due to the insufficient gain (*4*, *10*, *13*). Moreover, in a plasmon nanolaser, with increasing optical confinement, the momentum mismatch between a laser cavity mode and the outside free-space light increases rapidly, resulting in decreasing efficiency in light-matter interaction for practical applications (*14*).

On the other hand, complying with the electromagnetic boundary conditions (*15*), tight field confinement (e.g., down to 10-nm level) has recently been demonstrated in all-dielectric interfaces such as slot waveguides (*16*, *17*) and bowtie-shaped structures (*18–21*), indicating a possibility to surpass the field confinement of plasmonic nanostructures if the interface size can be further scaled down. Moreover, compared with plasmonic confinement, all-dielectric confinement has much lower optical loss and smaller photon momentum mismatch, which may circumvent the issues in plasmonic nanolasers such as heat-induced damage and low output efficiency.

Here we demonstrate a coupled nanowire pair (CNP) nanolaser that can circumvent the confinement-loss trade-off and offer a subnanometer field confinement ($\sim\lambda_0/1000$). We assemble two high-gain CdS nanowires into a CNP and form a slit-waveguide-based all-dielectric nanocavity. Pumped by 355-nm-wavelength laser pulses, we observe a low-threshold $TE_0$-like lasing mode around 520-nm wavelength supported by low-loss oscillation of polarized bound electrons surrounding a 1-nm-width slit, which offers a dominant peak with a field confinement down to 0.23 nm and a peak-to-background ratio of ~32.7 dB.

## Results

### Ultratight optical field confinement in a CNP

The structure of the nanolaser is schematically illustrated in Fig. 1A. The CNP, supported on a low-index $MgF_2$ substrate, is formed by a pair of CdS nanowires synthesized via a chemical vapor deposition method (*22*). The single-crystal wurtzite CdS nanowire has a uniform diameter (defined as the diagonal of its hexagonal cross-section), an atomic-level sidewall smoothness and a gain high enough (e.g., >1 $\mu m^{-1}$ around 500-nm wavelength of the bandgap transition (*6*)) for supporting lasing oscillation in a short (e.g., 10-$\mu m$ level) cavity (*23*). The two identical nanowires are cut from the same nanowire, assembled in close contact in parallel via micromanipulation, and then milled via focused ion beam to form a CNP with flat endfaces (Fig. 1B, see also Methods and Fig. S1). Since the measured surface undulation of the nanowire is ~1 nm (section S1), the slit between the vertex edges of the two nanowires can be approximately treated as a 1-nm-width slit (Fig. 1C) with a V-shaped index profile (inset of Fig. 1C).

Relying on the coupled oscillation of polarized bound electrons around its both sides (section S2), the slit can support a hybrid TE-like waveguiding mode at 520-nm wavelength (typical lasing wavelength of CdS nanowires (*24*)) with a central hotspot in the cross-section of the CNP (Fig. 1D). When the slit width is small enough (e.g., 1-nm scale), the TE-like mode can offer a central peak at the hotspot with ultratight field confinement and sufficiently large (e.g., >30 dB) peak-to-background intensity contrast (section S3). Figures 1E and 1F give the calculated full width at half maximum (FWHM) of the field intensity of the $TE_0$-like mode (the lowest-order and also the fundamental waveguiding mode), the 0.23-nm ($x$-axis) and 1.05-nm ($y$-axis) FWHMs offer a spot size down to ~ $5.5\times10^{-6}\lambda_0^2$ (Fig. 1G).

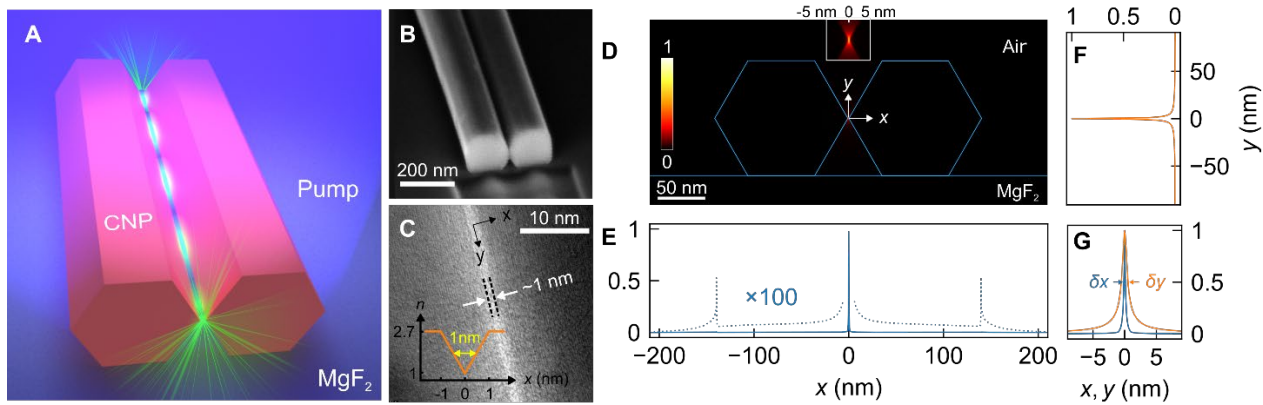

**Fig. 1. CNP for ultratight optical field confinement. (A)** Schematic illustration of a CNP-based nanolaser. **(B)** Scanning electron microscope (SEM) image of an as-fabricated CNP with a flat endface. **(C)** Transmission electron microscope image of a 1-nm-width slit formed between two nanowires. Inset, index profile around the slit. **(D)** Cross-sectional electric field intensity of a $TE_0$-like mode supported by a CNP with a nanowire diameter of 140 nm and a slit width of 1 nm. Inset,

close-up view of the field intensity around the slit. **(E and F)** Intensity profile of the field along horizontal (*x*-axis) and vertical (*y*-axis) directions in (D) across the slit center. For better clarity, a 100× profile is also plotted as dotted lines in (E). **(G)** Close-up view of the intensity profile in (E) and (F) around the slit center. The wavelength in the calculation is 520 nm.

Apart from the central peak, the $TE_0$-like mode has a low-intensity background field that is largely flat inside the CdS nanowire and faded out quickly beyond the nanowire's farmost edges (dotted line in Fig. 1E). Unlike a plasmonic mode that can be wholly sub-diffraction confined, here the $TE_0$-like mode as a whole is diffraction-limited. However, the subnanometer-level low-loss field confinement is far beyond the reach of a plasmonic mode (section S4). A certain background field is beneficial to bestow the central peak with high intensity and slow decay while maintaining an extreme optical confinement. Also, when interfacing with free-space photons, the momentum mismatch of an all-dielectric mode shown here is much smaller than that of an ultra-confined plasmonic mode, making it much more efficient for light-matter interaction in the near field outside the cavity (section S4).

It is worth to mention that, within a ~0.0045% of the total mode area, the ultra-confined central peak concentrates ~0.10% of the total mode power (section S5). Calculated fraction of the mode power contained inside the CdS nanowire is ~50.6% (section S5), contributing to a confinement factor $\Gamma$ (the overlap between the mode field and the gain region, see equation (2) and Fig. S7A) of ~0.68 for the CNP, which is favorable for reducing the lasing threshold.

**Laser modeling of a CdS CNP cavity**
To explore the lasing modes in the CNP, we firstly calculate its guiding modes at 520-nm wavelength with a slit width *w* of 1 nm. When the nanowire diameter *d* is less than 155 nm, the CNP supports only $TE_0$- and $TM_0$-like modes (Fig. 2A and S8A). The $TE_0$-like mode is mainly horizontally polarized (Fig. 2B), while the $TM_0$-like mode is basically vertically polarized (Fig. 2C). Theoretically, based on the cutoff diameter (108 and 118 nm for $TE_0$- and $TM_0$-like modes, respectively), a CNP can supports only a $TE_0$-like mode when 108 nm < *d* < 118 nm. However, experimentally, in a nanowire with such a small diameter, the substrate-induced leakage loss and the lasing threshold increases drastically with decreasing nanowire diameter (*6*, *25*). To facilitate the experimental realization, here we use nanowires with diameters slightly less than the cutoff of the third mode (i.e., ~155 nm of the $TE_1$-like mode, see also Fig. S8A), since the first two modes ($TE_0$- and $TM_0$-like) can be readily identified by polarizations.

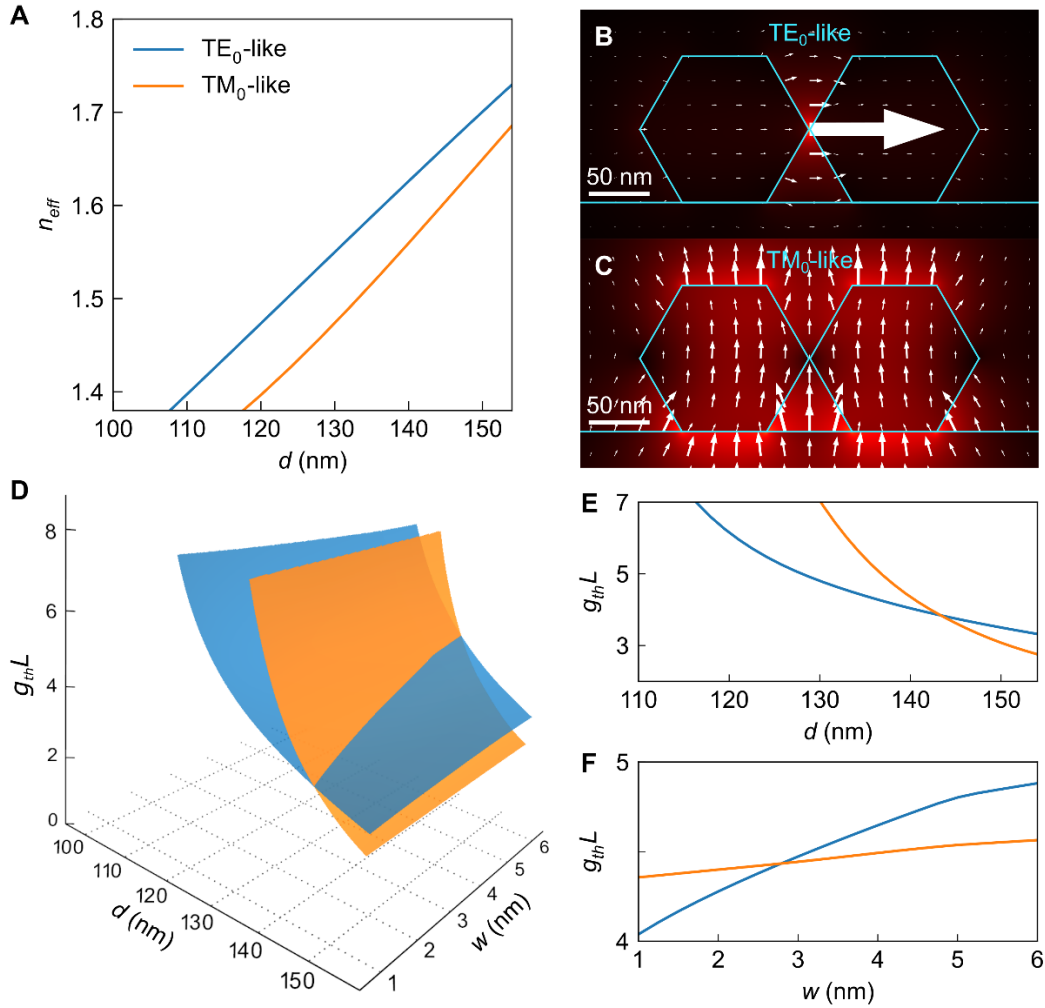

**Fig. 2. Laser modeling of a CdS CNP cavity.** (**A**) Effective refractive index $n_{eff}$ of the $TE_0$- (blue) and $TM_0$-like (orange) modes in a CNP with $w = 1$ nm. (**B** and **C**) Electric field vectors of the $TE_0$- and $TM_0$-like modes in a CNP with $d = 140$ nm and $w = 1$ nm. The orientation and size of the white arrow indicate the polarization and amplitude of the local field. (**D**) Threshold gain $g_{th}L$ of $TE_0$- (blue) and $TM_0$-like (orange) modes with varying $d$ and $w$. (**E** and **F**) Projection of $g_{th}L$ in (D) in the plane of (E) $w = 1$ nm and (F) $d = 140$ nm, respectively. The wavelength in the calculation is 520 nm.

In a Fabry-Perot-type CNP cavity, the lasing threshold gain $g_{th}$ can be obtained as $g_{th}L = 1/(\Gamma \ln R)$ (*26*), where $L$ is the nanowire length and $R$ the single-trip reflectivity. Calculated $g_{th}L$ (Fig. 2D) shows that, lasing thresholds of both modes increase monotonously with decreasing $d$ and increasing $w$, which is reasonable as smaller $d$ offers lower effective gain and high round-trip loss, and increasing $w$ decreases $\Gamma$. In particular, compared with the $TM_0$-like mode, the $TE_0$-like mode has a lower threshold with $d < 143$ nm when $w = 1$ nm (Fig. 2E), or $w < 2.8$ nm when $d = 140$ nm (Fig. 2F).

**Experimental lasing behavior in a CdS CNP**

To investigate the lasing behavior experimentally, we use a focused beam of 355-nm-wavelength laser pulses (3.5-ns pulse width and 1-kHz repetition rate) to pump a CNP (Fig. 3A) with $d = 140$ nm, $w = 1$ nm and $L = 23$ μm (Figs. 3B and 3C, see also Methods and Fig. S9A). With increasing pump density, the evolution of the CNP emission can be categorized into four stages (Fig. 3D): (1) under a low pump density (e.g., 24.0 kW/cm$^2$), the output endface of the CNP is a dim green-color spot in optical micrograph (bottom left inset), corresponding to a 14-nm-bandwidth photoluminescence centered around 515-nm wavelength (Fig. S9B); (2) with increasing pump density, two evident peaks emerge with similar intensity and orthogonal polarizations (e.g., at 36.5 kW/cm$^2$, see also Fig. S10A to E), corresponding to the horizontally polarized ($P_{//}$) TE$_0$-like mode and vertically polarized ($P_\perp$) TM$_0$-like mode, respectively; (3) when the pump density is above a certain threshold (36.5 kW/cm$^2$, see right inset), the intensity of the TE$_0$-like mode (around 519-nm wavelength) increases rapidly, accompanying with evident linewidth narrowing (Fig. S10F), while the TM$_0$-like mode (around 517-nm wavelength) remains low, showing that the lasing emission comes first from the TE$_0$-like mode. Meanwhile, the endface output is much brighter with clear interference fringes (top left inset), confirming the higher intensity and better coherence of the emission above the lasing threshold; (4) when the pump density is further increased (e.g., 43.7 kW/cm$^2$), the TM$_0$-like mode is also in lasing state (right inset).

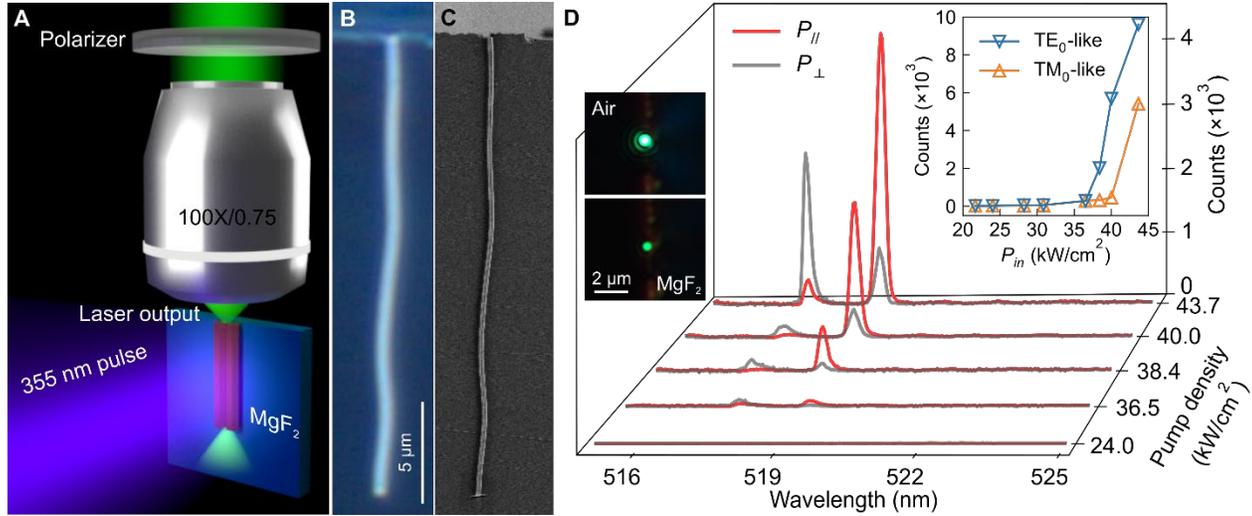

**Fig. 3. Experimental lasing behavior in a CdS CNP. (A)** Schematic of the setup. **(B and C)** Optical microscope and SEM images of a CdS CNP with $d$ = 140 nm and $w$ = 1 nm. **(D)** Measured lasing spectra with horizontal ($P_{//}$, red) and vertical ($P_\perp$, gray) polarizations obtained from the CNP endface. Left insets, optical microscope images of endface output under low (bottom) and high (upper) pump densities. Right inset, threshold measurement of the TE$_0$-like ($P_{//}$) and TM$_0$-like ($P_\perp$) modes.

The measured lasing threshold (right inset, Fig. 3D) of the TE$_0$-like mode is evidently lower than that of the TM$_0$-like mode (36.5 versus 40.0 kW/cm$^2$), which agrees well with theoretical calculations (Fig. 2D and 2E), and confirms that the CNP laser runs in single TE$_0$-like mode with pump density between the two thresholds. The polarization of lasing emission is also measured (Fig. 4A), in which the angle-dependent intensity of the single- and dual-mode state agrees very well with theoretical calculations.

**Lasing mode with extreme field confinement**

Since the subnanometer-level spatial confinement is far beyond the resolution of current optical imaging techniques (e.g., 20 nm of a near-field scanning optical microscope (*21*, *27*)), here we calculate the modal profile of the TE$_0$-like mode using a finite element method, which remains valid for all-dielectric structures with 1-nm-level feature sizes (*28*). Figure 4B shows the intensity profile of the TE$_0$-like lasing mode retrieved on the endface plane. The prominent central peak, with 0.23-nm field confinement, is about 180 times (22.6 dB) and 1800 times (32.7 dB) higher in field intensity ($|E|^2$) than the highest side peaks and overall background (see Fig. S4), respectively. The near-field intensity evolution of the light emitting into free space is also investigated (Fig. 4C). As it escapes out of the CNP endface, the field of the central peak spreads out quickly with decreasing intensity (inset of Fig. 4C), a similar behavior as light transmission through other deep-subwavelength apertures (*29*, *30*), which can be interpreted as an inevitable result of the Heisenberg uncertainty relation between the photon momentum and spatial confinement (see section S4). However, since the majority of the mode power is distributed in the diffraction-limited background field, the effective index of the $TE_0$-like mode is 1.63 (Fig. 2A), leading to a transmissivity as high as 94% (see Fig. S7B, effective reflectivity ~0.06), much higher than those of deep-subwavelength apertures even using plasmonic background fields for momentum compensation (*31*). Moreover, the relatively slow intensity decay of the output field is particularly desirable for near-field light-matter interaction for nanoscale objects (see section S4). In addition, the calculated field gradient of the central peak is about $6.3 \times 10^{19}$ $V^2/m^3$ (1-nW power, see section S6), which is promising for nanoscale manipulation using optical gradient force (e.g., cold atom manipulation (*32*)). Moreover, without the tunneling-induced quenching of field enhancement (e.g., in plasmonic structures with subnanometer gaps (*33*)), the slit width of the all-dielectric nanolaser may be further scaled down to obtain laser field with tighter field confinement.

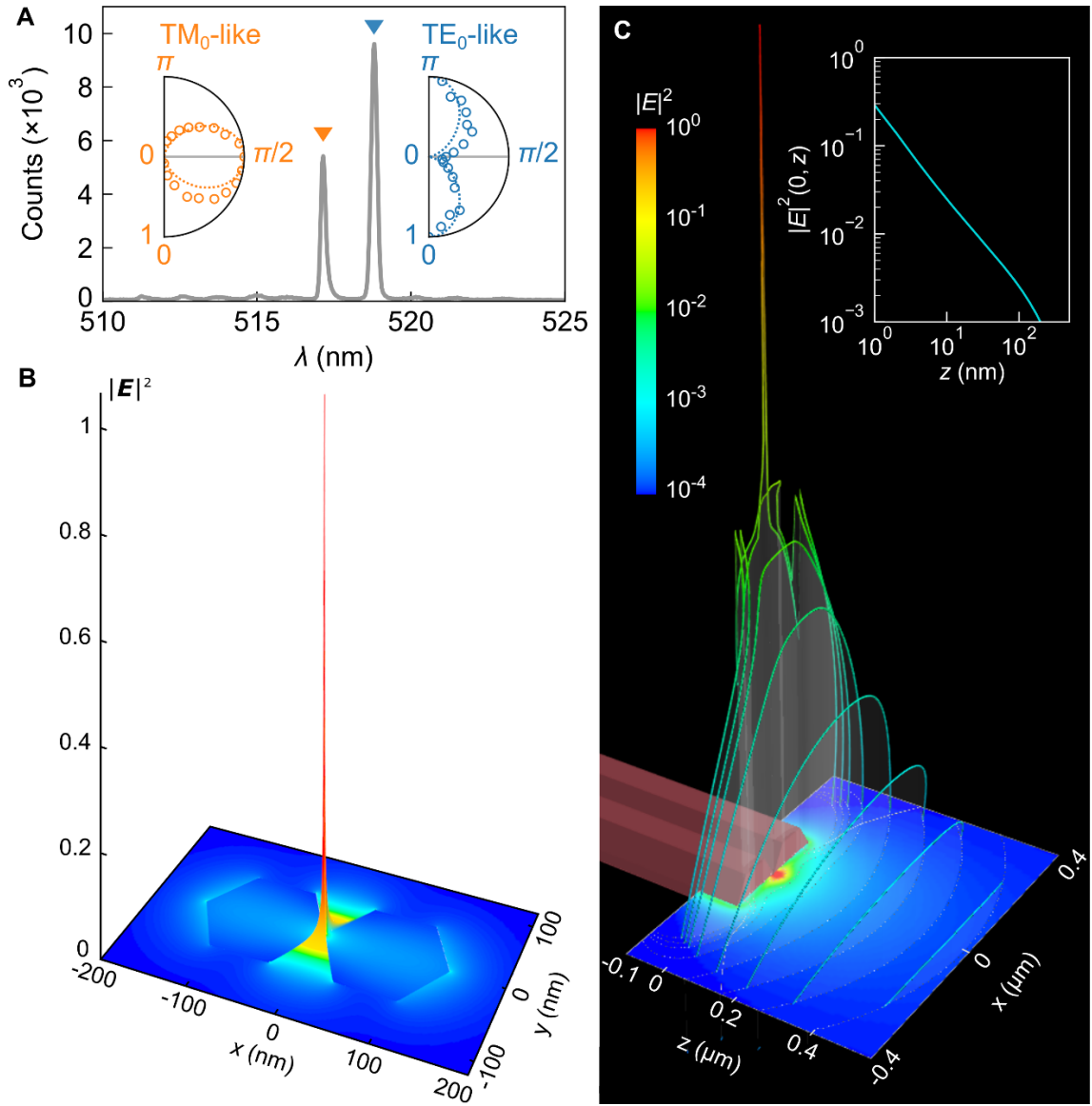

**Fig. 4. Field characterization of the $TE_0$-like lasing mode.** (**A**) Lasing spectrum at pump density of 43.7 $kW/cm^2$. Insets, calculated (dotted lines) and measured (open circles) angle-dependent spectra intensity of $TE_0$- (blue) and $TM_0$-like (orange) modes. (**B**) Normalized electric field intensity $|\textbf{\textit{E}}|^2$ of the $TE_0$-like mode at the endface of the CNP. (**C**) $|\textbf{\textit{E}}|^2$ of the $TE_0$-like mode on the middle cut plane of a CNP and its evolution (in logarithmic scale) along the isophase of $E_x$ with

varying distances away from the endface. Inset, the evolution of $|E|^2(0, z)$ in log-log scale.

**Discussion**

Overall, we have demonstrated an all-dielectric CNP nanolaser with subnanometer-level field confinement. Compared with existing nanolasers, this laser can offer not only much tighter field confinement, but also low-loss operation and high-efficiency output, which may open an avenue for light-matter interaction on much lower dimensions, e.g., efficiently interfacing photons with single molecules or atoms with high spatial selectivity. With its unparalleled field confinement, the nanolaser shown here may inspire a new category of nanolasers and push the limit of laser-based science and technology beyond the nanometer scale.

**Materials and Methods**
**Sample preparation.**
CdS nanowires were manipulated by tapered fiber probes mounted on 3-dimensional moving stages under an optical microscope (*34*). Firstly, using two tapered fiber probes for micromanipulation under an optical microscope, a nanowire was picked up from the collection silicon wafer (*22*) and transferred onto the surface of a MgF$_2$ substrate. Secondly, following a bend-to-fracture process (*35*), the nanowire was cut into two at the center using a fiber probe. Then, using fiber probes for micromanipulation, the two nanowires were pushed to one edge of the MgF$_2$ substrate, and assembled to contact in parallel to form a CNP, with its length perpendicular to the edge of the substrate. Finally, using focused ion beam (FIB, Auriga 40, Carl Zeiss) milling process with a milling current of 1 pA, both ends of the CNP were cut with flat endfaces to form a CNP cavity based on the endface reflection. Details of the step-by-step preparation process can be found in Fig. S1.
**Calculation of lasing threshold gain of the waveguiding modes in a CNP.**
Relying on endface reflection, a CNP can be treated as a Fabry-Perot cavity, and the lasing threshold gain ($g_{th}$) of its waveguiding modes can thus be obtained by (*26*)

$$\Gamma g_{th} = \frac{1}{L} \ln \left(\frac{1}{R}\right), \tag{1}$$

where $\Gamma$ is the confinement factor, $L$ the cavity length and $R$ the endface reflectivity. As a dielectric waveguide, here the $\Gamma$ can be obtained as (*36*)

$$\Gamma = \frac{c\epsilon_0 n_{NW} \iint_{CNP} \frac{1}{2}|E|^2 dxdy}{\iint P_z \, dxdy}, \tag{2}$$

where $c$ is the light velocity in vacuum, $\epsilon_0$ the vacuum permittivity constant, $n_{NW}$ the refractive index of CdS, and $P_z$ the $z$ component of the Poynting vector of the waveguiding modes. The endface reflectivity is obtained using finite-difference time-domain (FDTD) simulation (*37, 38*). Based on the above-mentioned results, we obtained the lasing threshold gain $g_{th}$ of TE$_0$- and TM$_0$-like modes in a CNP, as shown in Figs. 2D-F in main text.
**Experimental details.**
The experiments are performed at room temperature. A beam of 355-nm-wavelength laser pulses (FTS355-Q4, CryLaS GmbH, 3.5-ns pulse width and 1-kHz repetition rate) is loosely focused on the CNP perpendicular to its length (Fig. S9A, see also Fig. 3A). As shown in Fig. S9B, the CNP is located around the center of the 500-μm-diameter focusing spot, which covers the whole length of the CNP with nearly uniform intensity.

The CNP used here (assembled with nanowire diameter $d$ of 140 nm, slit width $w$ of 1 nm) is about 23 μm in length after FIB milling (Fig. 3, B and C). To facilitate FIB milling and SEM imaging, the MgF$_2$ substrate has been sputtered with a thin layer of Au film except a small area

(about 30 μm × 50 μm) surrounding the CNP (Fig. S9C). One end of the CNP is placed perpendicular to the edge of the substrate (see also Fig. S1F) for facilitating the collection of endface emission. The output from the endface at the substrate edge is collected by an objective lens (Objective LD EC Epiplan-Neofluar HD DIC M27, Zeiss, 100×, NA = 0.75), routed through a polarizer, and then directed to a CCD camera and a spectrometer (iHR550, HORIBA, 0.035 nm spectral resolution), as shown in Fig. S9A.

**Acknowledgments**
The authors thank Y. X. Gao for helpful discussion.
**Funding:** This work is supported by the National Key Research and Development Project of China (2018YFB2200404), the National Natural Science Foundation of China (61635009 and 11527901), and the Fundamental Research Funds for the Central Universities.
**Author contributions:** H.W. conceived the idea, performed the calculation and experiments. H.W. and L.M.T. wrote the manuscript. P.Z.X. provided the CdS nanowires and participated in the experiment. X.G., P.W. and L.M.T. supervised the project.
**Competing Interests:** The authors declare no competing interests.
**Data and materials availability:** All data needed to evaluate the conclusions in the paper are present in the paper and/or the Supplementary Materials.

# Supplementary Text

## S1. Geometry of CdS nanowires and CNPs

Figures S2A and S2B present the SEM images of typical individual single-crystal CdS nanowires synthesized via chemical vapor deposition (*1*), the hexagonal cross sections are clearly seen. Figures S3C and D give high-resolution transmission electron microscopy (HR-TEM, JEM-2100, JEOL. The operation voltage is 200 kV) images of a side edge of a 130-nm-diameter nanowire, showing excellent sidewall smoothness. By measuring the undulation amplitude of the nanowire surface in Fig. S3D, we obtain a peak-to-valley undulation of ~1 nm.

The slit size in a CNP was also investigated under an HR-TEM. Figure S3E shows a typical CNP with nanowire diameter of 150 nm. When a beam of accelerated electrons was focused and incident perpendicularly onto the slit (Fig. S4F), it partially penetrates through the bowtie structure (Fig. S4G) with position-dependent transmission flux (*2*)

$$I(x) = I_0 exp(-Qd(x)), \quad (S1)$$

where $I_0$ is the initial flux of the beam, $Q$ is the scattering cross section, and $d(x)$ is the thickness of sample along the traveling path of the electrons. Thus, from the HR-TEM image (Fig. S4H), we obtain $I(x)$ at different $z$ (Fig. S4I), which agree well with that predicted by equation (S1) excepting the central area (i.e., the slit), where the edge material is randomly distributed due to the sub-nm surface undulation. By subtracting the fit value (dashed lines) from the measured $I(x)$, we obtain a slit width from 0.5 to 1.1 nm (Fig. S4J), i.e., on 1-nm level. Therefore, it is reasonable to assume that the slit width is 1 nm in this work.

## S2. All-dielectric field confinement around the slit in a CNP via polarized bound electrons

When the $TE_0$-like mode is propagated around the slit in a CNP, the ultratight field confinement is realized via polarized bound electrons. Figure S3A presents the density distribution of polarization $\hat{n} \cdot P$, where $\hat{n}$ is the unit direction vector and $P$ the dielectric polarization, of the polarized bound electrons (*3*). The high density of polarization (oscillating at optical frequency) in the two opposite edges (the opposite vertexes in the cross-sectional view in Fig. S3B) offers ultratight confinement of the optical field waveguided along the slit.

## S3. Peak-to-background ratio of the confined field in the $TE_0$-like mode with different slit widths

The peak-to-background ratio ($R_{P1/B}$ and $R_{P2/B}$) of $TE_0$-like mode is calculated as $R_{Pi/B} = 10\log(I_{Pi}/I_B)$, $i$=1, 2,… where $I_{P1}$ is the field intensity of the dominant central peak, $I_{P2}$ the field intensity of the 2nd-largest peak (i.e., the largest side peak) and $I_B$ the averaged field intensity over the mode area, defined as $I_B = \frac{\iint_\Omega |E|^2 dxdy}{\Omega}$, where $\Omega$ is the mode area defined in section VII.

The calculated results are shown in Fig. S4. For comparison, mode profiles with slit widths of 1 nm, 10 nm and 50 nm are depicted, clearly showing the increasing $R_{P1/B}$ with decreasing slit width.

# S4. Comparison of spatial confinement limit of output field of different kind of nanolaser and Heisenberg uncertainty relation of spatial confinement and transmission for a coherent light field

So far, although there is no experimental study on optical transmission through an aperture with size far below 10 nm (e.g., a 1-nm-diameter circular aperture made on a thin metal film with ideal conductivity), the intensity profile of the transmitted light field can be theoretically estimated using Heisenberg uncertainty principle (*4*). According to the Heisenberg uncertainty principle of photons (*5*, *6*),

$$\Delta x \cdot \Delta p \geq \alpha \hbar, \tag{S2}$$

where the spatial uncertainty $\Delta x$ is approximately the size of spatial confinement, and $\Delta p$ is the momentum uncertainty of the photons. For a coherent light field (i.e., laser field in this work), $\alpha = \frac{3}{2}\sqrt{1 + \frac{9}{4}\sqrt{2}}$.

Considering the photon momentum conservation before and after transmitting the aperture (Fig. S5C),

$$|\mathbf{k_0}|^2 = |\mathbf{k}|^2 = k_x^2 + k_y^2 + k_z^2, \tag{S3}$$

where $\mathbf{k_0}$ and $\mathbf{k} = (k_x, k_y, k_z)$ are the wave vector of a photon before and after transmitting the aperture, respectively. Thus, we have

$$k_z = \sqrt{k_0^2 - k_x^2 - k_y^2}. \tag{S4}$$

For a photon, the momentum $p = \hbar k$, from equation (S2), we have

$$\Delta x \cdot \Delta k_x \geq \alpha \quad \text{and} \quad \Delta y \cdot \Delta k_y \geq \alpha. \tag{S5}$$

In a tiny aperture with diameter $D$ much smaller than the wavelength, $\Delta x \sim \Delta y \sim D$, $\Delta k_x \sim k_x$, $\Delta k_y \sim k_y$. Therefore, from equation (S2), we have

$$k_z = i\sqrt{2(\alpha/\Delta x)^2 - k_0^2} \tag{S6}$$

and

$$E(z) \propto \exp\left(-\sqrt{2(\alpha/D)^2 - k_0^2}\,z\right), \tag{S7}$$

which means that the transmitted field in propagating direction $z$ evolves with an intensity profile of $\exp\left(-2\sqrt{2(\alpha/D)^2 - k_0^2}\,z\right)$.

More practically, we assume the incident light is ideally focused to the optical diffraction limit (i.e., a beam size of $\lambda/2$) (*7*), and has the same power as the TE$_0$-like lasing cavity mode in the CNP nanolaser, we then compare the output optical field regarding the intensity and its decay along the $z$-direction in Fig. S5B-C, in which $D = 1$ nm, $\lambda = 520$ nm, and optical power of the incident light (or the TE$_0$-like lasing mode inside the CNP cavity) is assumed to be 1 nW, which is a reachable value for nanowire lasers operated slightly beyond the lasing threshold (for reference, the maximum output power of a nanowire laser can go up to 60 μW) (*8*). It should be noticed that, in the case of the 1-nm-slit CNP nanolaser, we count only the intensity of the central peak (i.e., 0.1% of the total mode power). It is clearly seen that, the field intensity of the laser peak output from a CNP nanolaser is much higher than that of the transmitted light through a 1-nm aperture (~7 orders of magnitude higher). Moreover, when emitting into the free space, the decay of the output field intensity of the CNP nanolaser is much slower than that of the 1-nm aperture, which is particularly desirable for light-matter interaction in near field for a variety of nanoscale objects such as single atoms, small molecules to large molecules. For reference, some typical nanoscale samples (a NH$_3$ molecular, glucose, quantum dot, Bovine serum albumin protein, DNA segment and virus) are also plotted in

Fig. S5D, with their typical sizes corresponding to the abscissa with arrow lines.

## S5. Calculation of effective mode area and fractional power in the central peak of the TE0-like mode in a CNP

The effective mode area is defined as the area of an region $\Sigma$ that satisfying:

a) its boundary $\partial\Sigma$ is the contour of $P_z$ of the mode, i.e., $P_z(\partial\Sigma)$ is constant;

b) 86.5% (that is $1 - 1/e^2$) of the total power is included inside the region, i.e.,

$$\frac{\iint_\Sigma P_z \, dxdy}{\iint P_z \, dxdy} = 1 - \frac{1}{e^2}. \tag{S8}$$

Figure S6A and B shows the region of the TE$_0$-like mode in the CNP at 520-nm wavelength with $d$ = 140 nm and $w$ = 1 nm. The effective mode area is calculated as around $0.50(\lambda/2)^2$ with $\lambda$ = 520 nm.

It is worth to mention that, in the cross-sectional plane shown in Fig. S6B, around the 4 inner vertexes of the 2 nanowires, $P_z$ could be very small or even negative (9).

Fractional power inside the solid CNP structure (i.e., the two nanowire) is calculated by integrating $P_z$ over the two separated hexagons (i.e., the cross-sections of the two nanowires) shown in Fig. S6C.

Fractional power in the central peak is calculated as,

$$\eta = \iint_\Omega P_z \, dxdy \, / \iint P_z \, dxdy, \tag{S9}$$

where the integral region $\Omega$ is a single closed area with $P_z(x,y) > P_z(0,0)/2$ as shown in Fig. S6D. Figure S6E shows calculated $\eta$ of TE$_0$-like mode with varying $w$ and $d$. The calculation is carried out at 520-nm wavelength.

## S6. Optical field gradient in the slit of CNP

Benefiting from its ultratight field confinement, the central peak of a TE$_0$-like mode in the CNP ($d$ = 140 nm, $w$ = 1 nm) can offer a field intensity gradient $\nabla \xi = \nabla |\boldsymbol{E}|^2$ up to ~$6.3 \times 10^{19}$ V$^2$/m$^3$ with a total power of merely 1 nW. Figure S11 presents the calculated field intensity gradient around the central peak on the output endface of the CNP nanolaser.

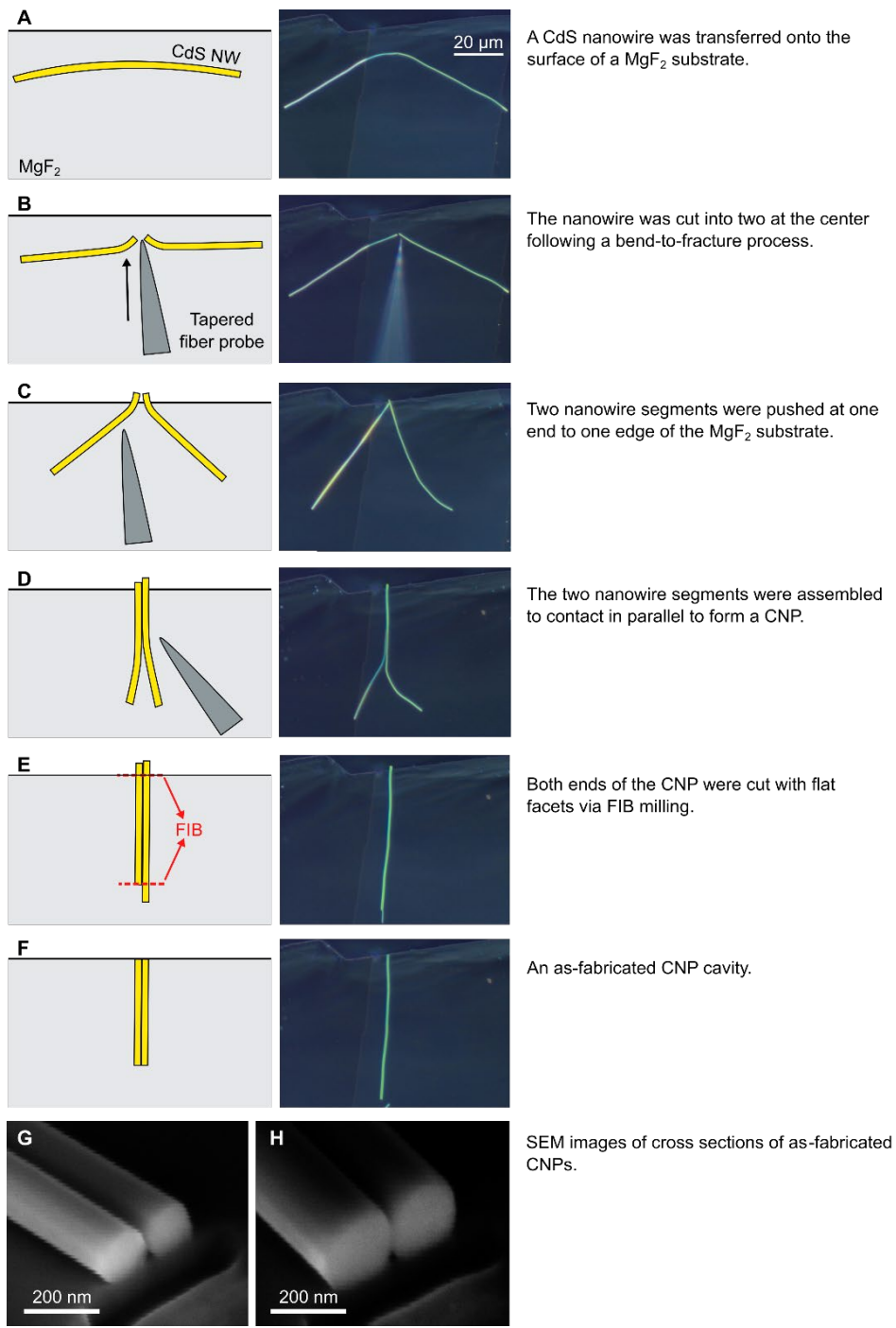

**Fig. S1. CNP preparation. (A** to **F)** Schematic illustration and optical microscope images in the CNP preparation. **(G** and **H)** SEM images of cross sections of as-fabricated CNPs with nanowire diameters of (G) 135 nm and (H) 200 nm, respectively.

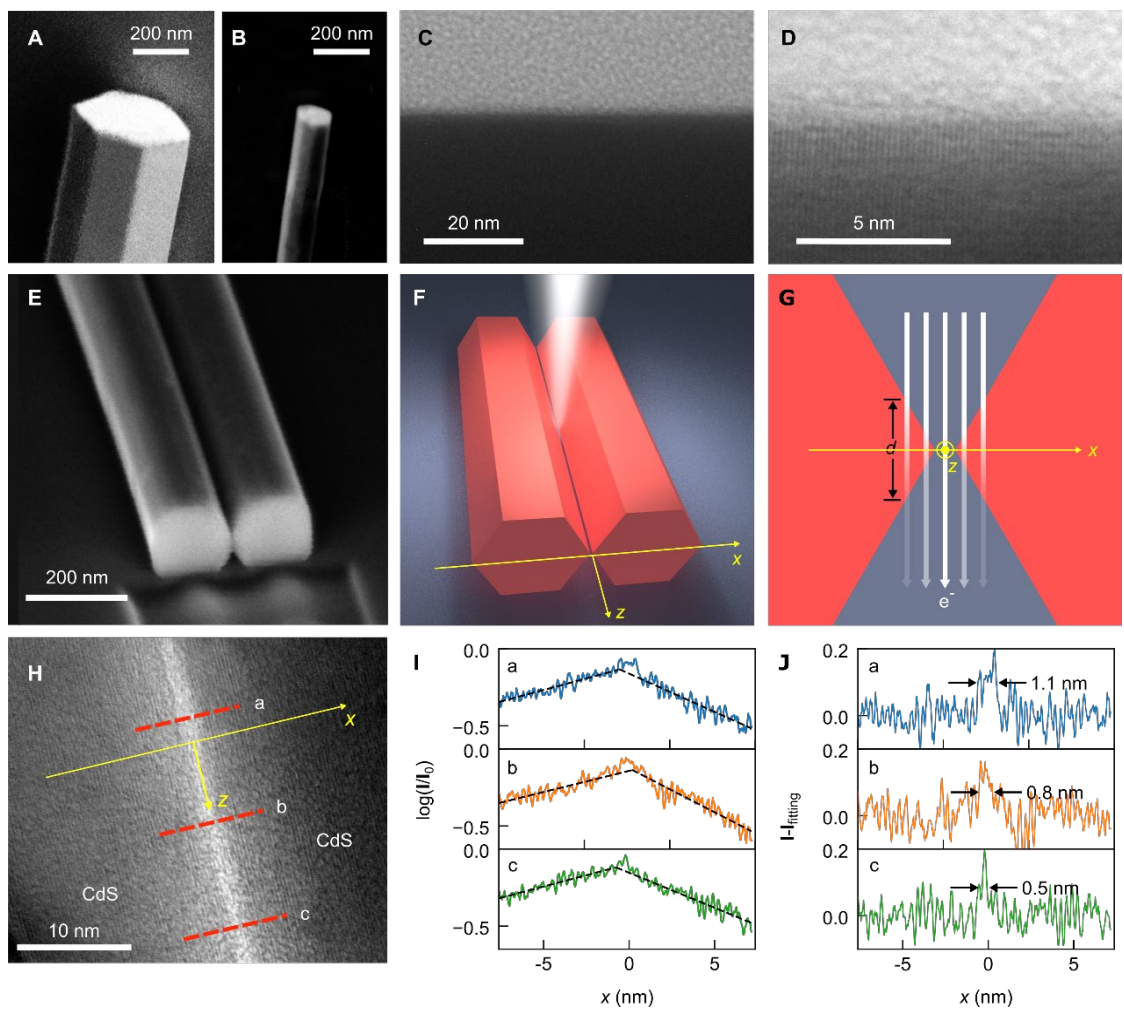

**Fig. S2. Electron microscopy study of CdS nanowires and CNPs.** (**A** and **B**) SEM images of endfaces of two individual CdS nanowires with diameters of (A) 420 nm and (B) 130 nm, respectively. The flat endfaces are obtained using a bend-to-fracture process. (**C** and **D**) HR-TEM images of a side edge of a 130-nm-diameter CdS nanowire. (**E**) HR-TEM image of a CNP with nanowire diameter of 150 nm. (**F**) Schematic illustration of a CNP with a vertically incident beam of focused electrons. (**G**) Schematic illustration of electron beam penetrating through the bowtie area surrounding the slit. (**H**) HR-TEM image of the central area surrounding the slit. The three lines marked by a-c are typical traces for measuring $I(x)$ at different $z$. (**I**) Measured $I(x)$ along lines a-c in (H), with approximately linear fitting lines (dashed black lines) in semi-logarithmic coordinates. (**J**), Slit widths obtained by subtracting the fitting value from the measured $I(x)$ in (E).

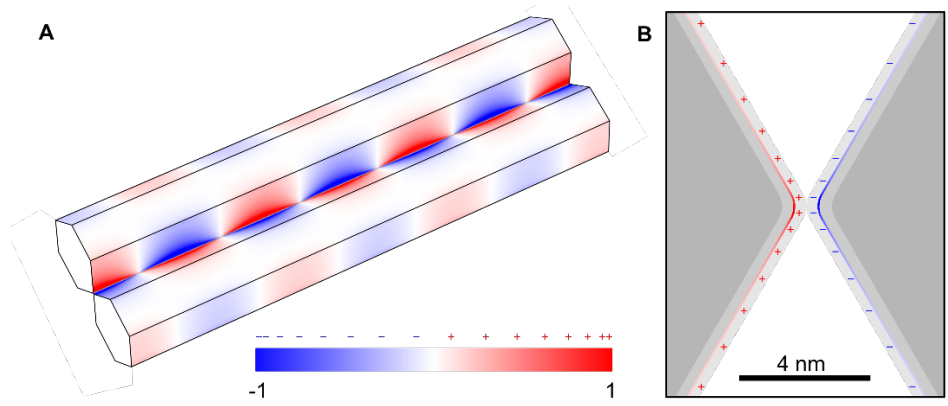

**Fig. S3.** **(A)** Schematic of density distribution of the polarized bound electrons in a CNP when a TE$_0$-like mode is waveguided through. **(B)** Cross-sectional density distribution of the polarization around the two opposite edges of nanowires in a CNP.

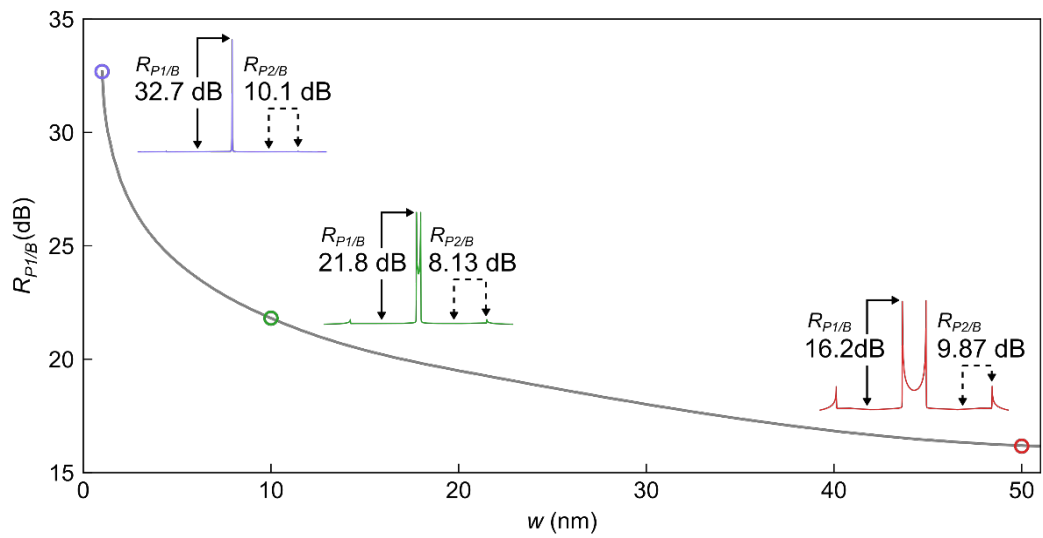

**Fig. S4.** Calculated peak-to-background ratio $R_{P1/B}$ of TE$_0$-like mode with varying slit width ($w$). Mode profiles at three typical slit widths of 1 nm (purple), 10 nm (green) and 50 nm (red) are also presented.

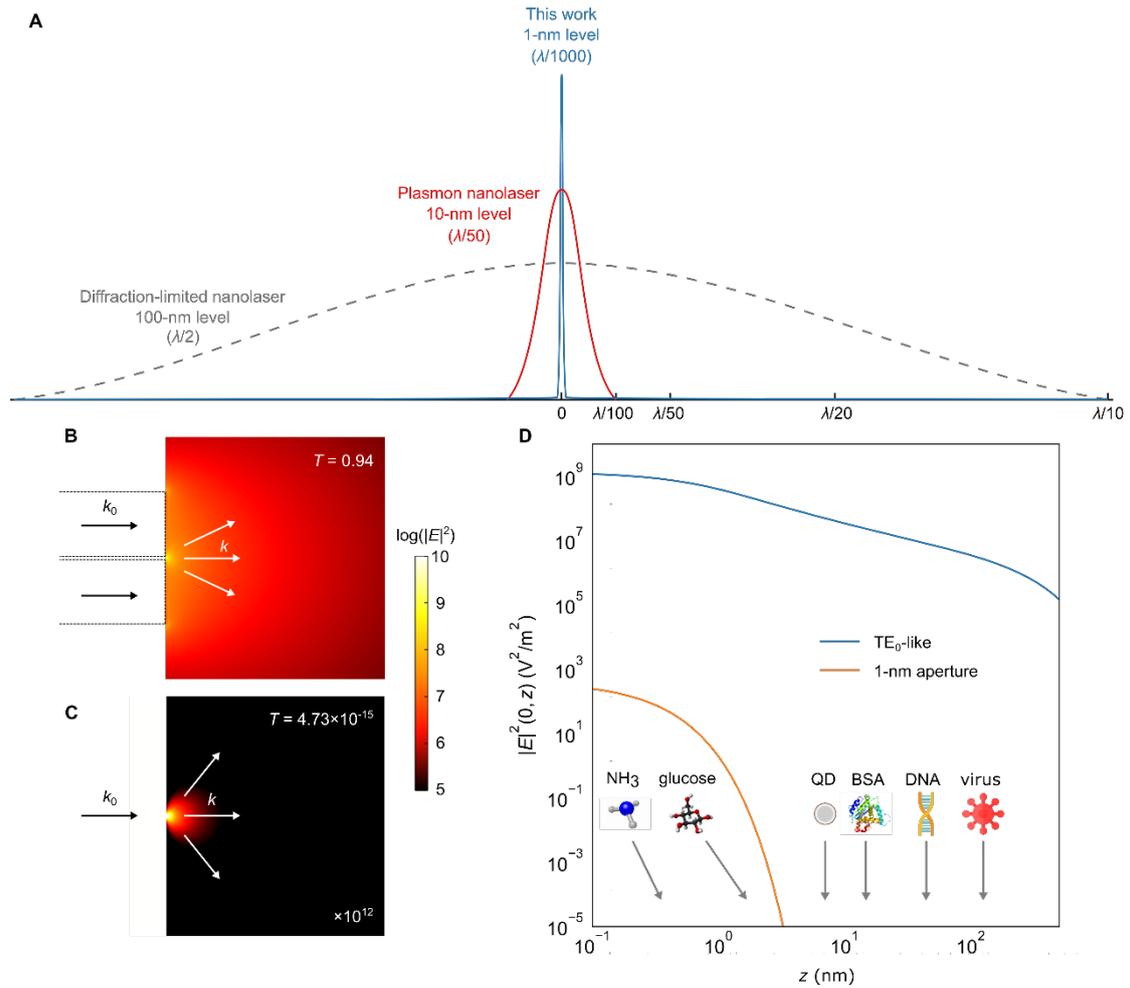

**Fig. S5. Comparison of spatial confinement limit and intensity evolution of output field along *z*-direction of lasing fields between a plasmon nanolaser and the CNP nanolaser in this work. (A)** Scale plot of the field confinement of a CNP nanolaser and a plasmon nanolaser, respectively. For reference, spatial confinement limit of a diffraction-limited nanolaser is also provided. **(B** and **C)** Logarithmic plot of electric field intensity of an optical field with a wave vector of $k_0$ (B) propagating out of a 1-nm-width slit of a CNP (dashed lines) with $d$ = 140 nm, and (C) passing through a 1-nm-dimeter circular aperture. The calculated transmissivity $T$ is also presented in each case. For better visualization, the output intensity in (C) is amplified by a factor of $10^{12}$. **(D)** Logarithmic plot of calculated output field intensity along *z*-direction of the $TE_0$-like mode (blue) and transmitted light through a 1-nm aperture (orange). The input field is 520 nm in wavelength and 1 nW in power. For reference, nanoscale samples (typical sizes) of a $NH_3$ molecule (~0.32 nm) (*10*), a glucose molecule (~1.5 nm) (*11*), a quantum dot (QD, ~7 nm), a Bovine serum albumin (BSA) protein (~14 nm) (*12*), a 0.1 kb DNA segment (~40 nm) and a virus (~120 nm) (*13*) are also provided.

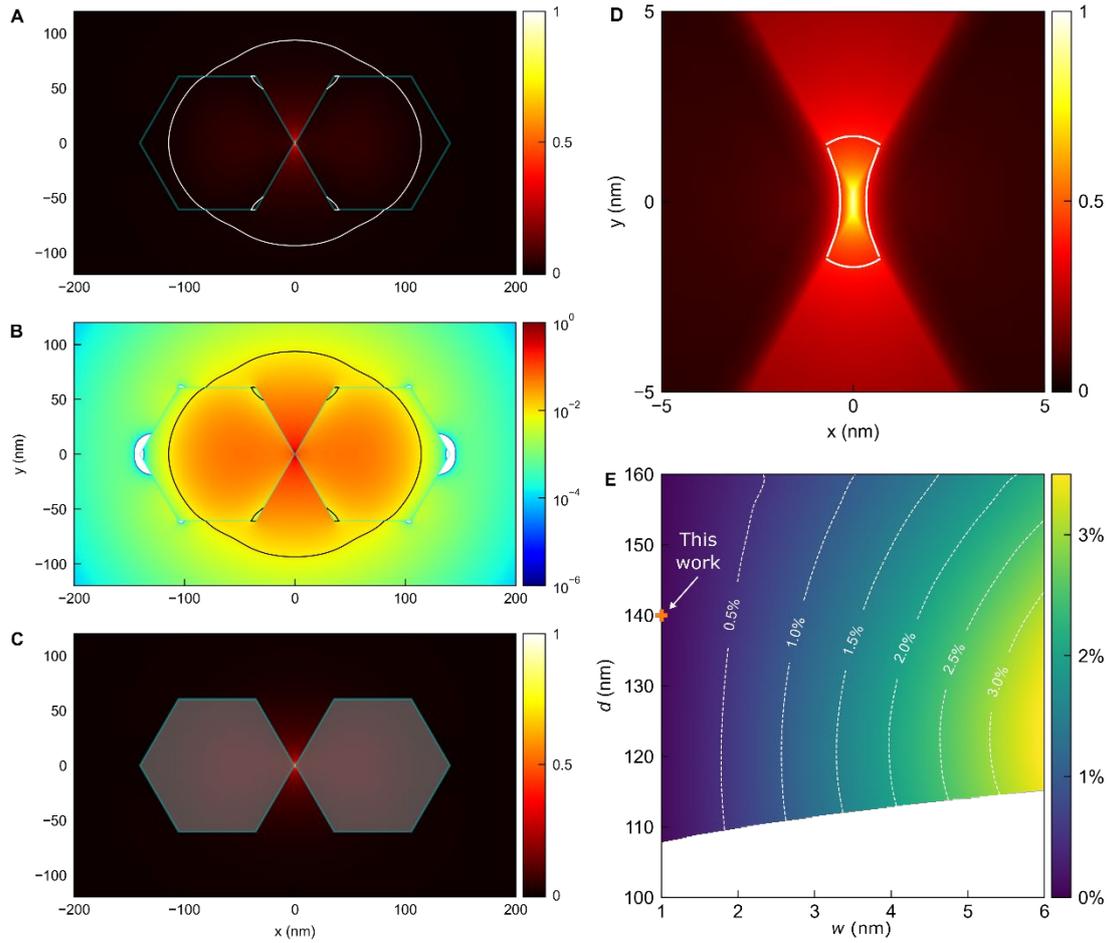

**Fig. S6. Calculation of effective mode area and fractional power in the central peak of the TE$_0$-like mode in a CNP. (A** and **B)** Boundary of effective mode area (white line in (A) or black line in (B)) of TE$_0$-like mode at 520-nm wavelength with $d$ = 140 nm and $w$ = 1 nm in linear and log scales, respectively. **(C)** Integration area (the two hexagons) for calculating the fractional power inside the CNP. **(D)** Integration area (inside the white contour line at level of $P(0,0)/2$) for calculating the fractional power inside the central peak. **(E)** Fractional power in the central peak over the total mode power. The parameters used in the experiment ($d$ = 140 nm, $w$ = 1 nm) is marked with a '+' sign.

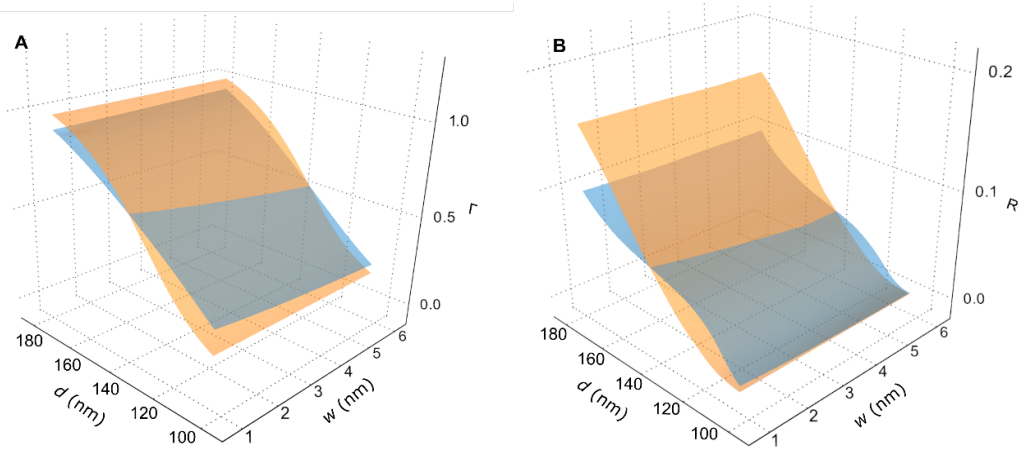

**Fig. S7.** Calculated **(A)** confinement factor $\Gamma$ and **(B)** endface reflectivity $R$ of TE$_0$- (blue) and TM$_0$-like (orange) modes with varying diameter $d$ and slit width $w$.

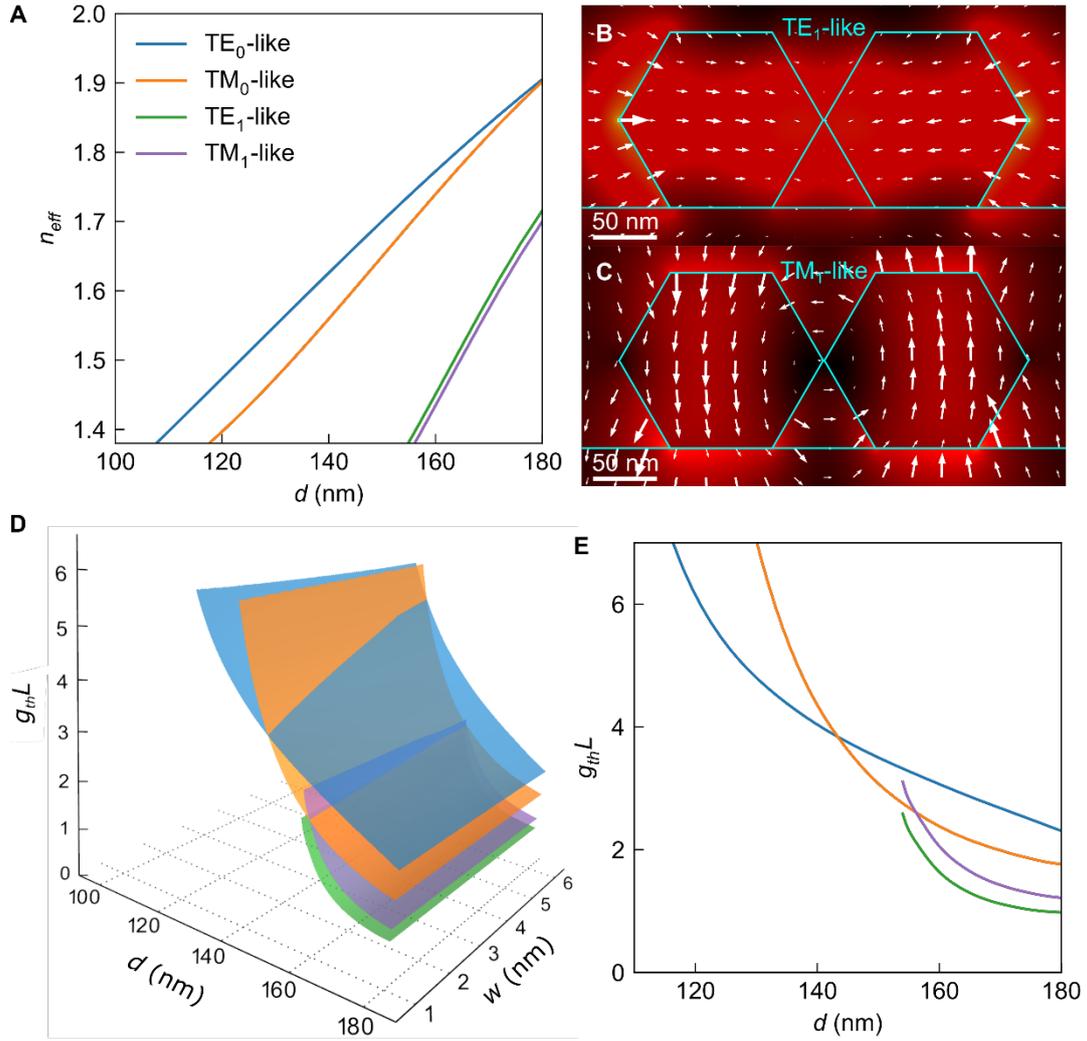

**Fig. S8.** Calculated parameters of the four lowest waveguiding modes in a CNP. **(A)** Effective refractive index $n_{eff}$. **(B and C)** Electric field vectors of $TE_1$- and $TM_1$-like mode in a CNP with $d$ = 160 nm. **(D)** Threshold gain $g_{th}L$ of $TE_0$- (blue), $TM_0$- (orange), $TE_1$- (green) and $TM_1$-like (purple) modes at 520-nm wavelength. **(E)** Dependence of $g_{th}L$ on $d$. The calculation in (A, B, C and E) is carried out at 520-nm wavelength with a constant $w$ of 1 nm.

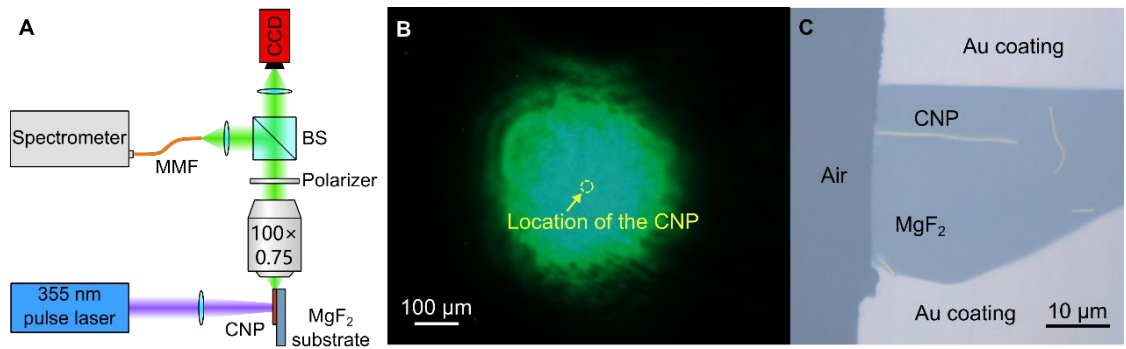

**Fig. S9. (A)** Schematic illustration of the experimental setup for optical pumping and lasing characterization of the CNP. BS: beam splitter; MMF: multimode fiber. **(B)** Optical microscope image of the focusing spot of the pump beam. The image is taken by placing a piece of CdS wafer at the focusing plane of the pump beam. A 25-μm-diameter dashed line circle at the center indicates the location of the CNP. **(C)** Optical microscope image of a CNP placed close to the edge of a MgF$_2$ substrate that is coated with a thin layer of Au film excepting the area supporting the CNP.

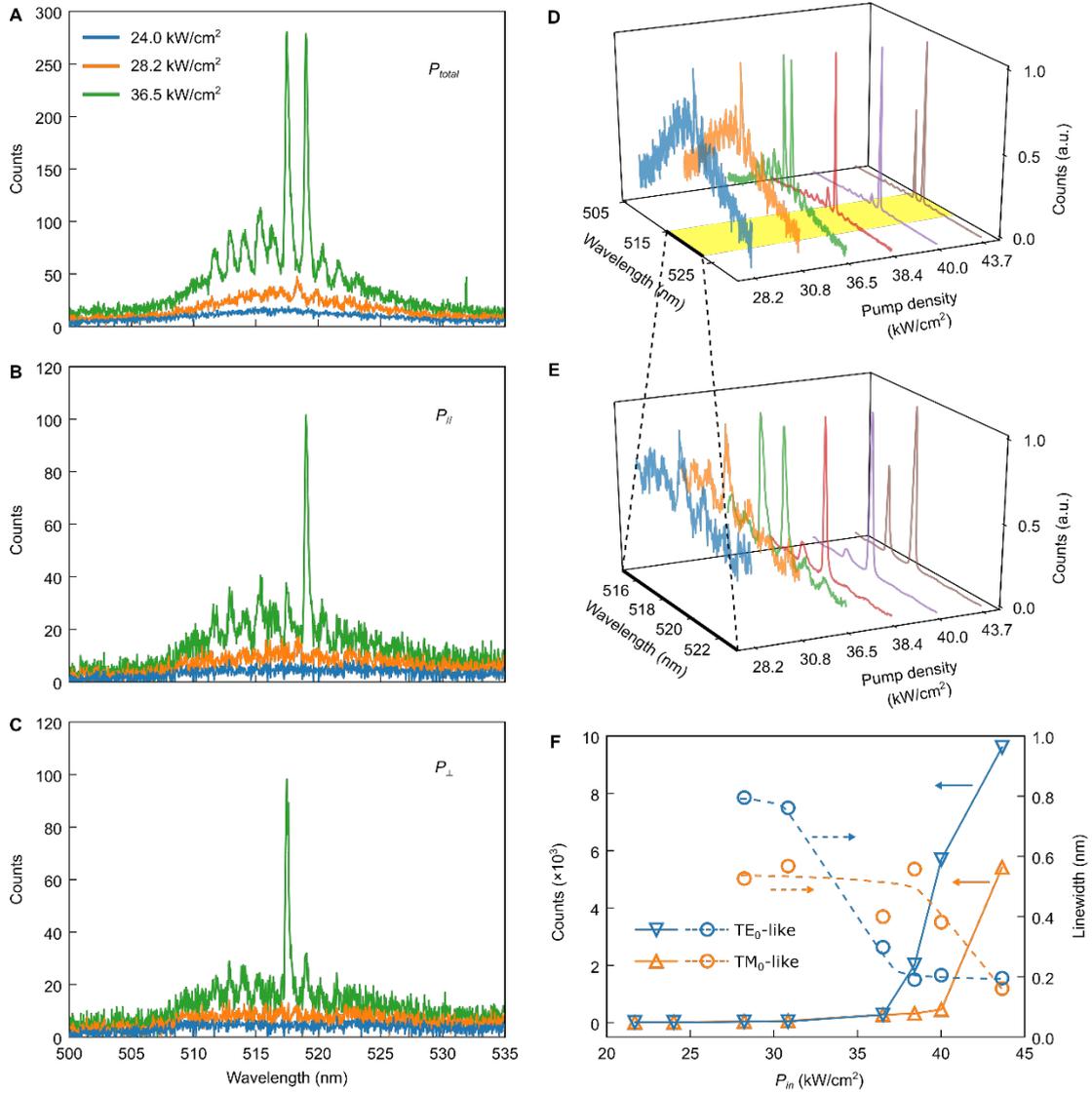

**Fig. S10. Details of the emission spectra of the CNP under different pump densities.** (**A** to **C**) Fluorescence spectra of the CNP collected with (A) no polarizer, (B) a $P_{//}$ (horizontal)-polarizer and (C) a $P_\perp$ (vertical)-polarizer, under pump densities of 24.0 kW/cm$^2$ (blue), 28.2 kW/cm$^2$ (orange) and 36.5 kW/cm$^2$ (green), respectively. (**D** and **E**) Measured normalized emission spectra of the CNP under different pump densities and close-up view of the spectra from 515 nm to 523 nm for better clarity. (**F**) Emission intensity (triangle, solid line) and linewidth (circle, dashed line) of the $TE_0$- (blue) and $TM_0$-like (orange) modes in the CNP.

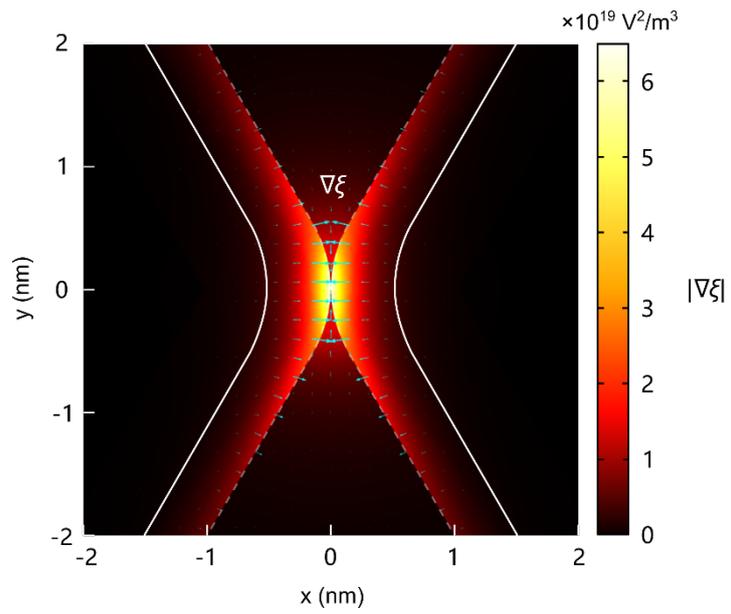

**Fig. S11.** Calculated field intensity gradient $\nabla\xi$ around the central peak of the $TE_0$-like mode on the output endface of the CNP nanolaser, with $d$ = 140 nm, $w$ = 1 nm, and a 1-nW mode power inside the cavity.